\documentclass[prd,nofootinbib]{revtex4}
\usepackage{amsmath}
\usepackage{graphicx}
\usepackage{amssymb}

\begin{document}

\title{Accretion processes in magnetically and tidally perturbed
Schwarzschild black holes }
\author{Zolt\'{a}n Kov\'{a}cs$^{1\dag }$, L\'{a}szl\'{o} \'{A}rp\'{a}d
Gergely$^{2,3\ddag }$, M\'{a}ty\'{a}s Vas\'{u}th$^{4\ast }$}
\affiliation{$^{1}$ Department of Physics and Center for Theoretical and Computational
Physics, The University of Hong Kong, Pok Fu Lam Road, Hong Kong\\
$^{2}$ ~Department of Theoretical Physics, University of Szeged, Tisza Lajos
krt. 84-86, Szeged 6720, Hungary\\
$^{3}~$ Department of Experimental Physics, University of Szeged, 6720
Szeged, D\'{o}m t\'{e}r 9, Hungary\\
$^{4}$ KFKI Research Institute for Particle and Nuclear Physics, Budapest
114, P.O.Box 49, H-1525 Hungary \\
$^{\dag }$zkovacs@hku.hk; $^{\ddag }$gergely@physx.u-szeged.hu; $^{\ast }$%
vasuth@rmki.kfki.hu}

\begin{abstract}
We study\ the accretion process in the region of the Preston-Poisson
space-time describing a Schwarzschild black hole perturbed by asymptotically
uniform magnetic field and axisymmetric tidal structures. We find that the
accretion disk shrinks and the marginally stable orbit shifts towards the
black hole with the perturbation. The radiation intensity of the accretion
disk increases, while the radius where radiation is maximal remains
unchanged. The spectrum is blue-shifted. Finally, the conversion efficiency
of accreting mass into radiation is decreased by both the magnetic and the
tidal perturbations.
\end{abstract}

\date{\today }
\maketitle

\section{Introduction}

As observational data on the astrophysical properties of the accretion disks
around black holes and other compact objects is accumulated, the study of
the accretion mechanism driven by these objects has become an important
research topic. The first and simplest theoretical model of the accretion
disks was constructed by imposing strong simplifications on the dynamics and
geometrical properties of the disk \cite{ShSu73,NoTh73}. In this so-called
steady-state thin accretion disk model a geometrically thin but optically
thick disk was considered in a hydrodynamic approximation by neglecting any
magnetic fields in the environment of the black hole and the disk. In their
early analysis, Novikov and Thorne \cite{NoTh73} modeled accreting matter as
a rotating fluid. This hydrodynamic approximation also holds in the presence
of a magnetic field as long as the deviation from geodesics of the photon
trajectories is less than the Larmor radius (which in turn is small compared
to the Schwarzschild radius). However, the discovery of the Blandford-Znajek
mechanism - describing how rotational energy can be extracted from a black
hole via magnetic field lines emanating from its event horizon - indicated
that magnetic fields can have a considerable effect both on the evolution of
the Kerr black holes and on the accretion processes feeding the black hole
with mass energy \cite{BZ77}. Later on magnetosphere models were introduced
for both static and rotating black holes, which allowed the study of the
both the effects of the flux lines connecting the black hole to the
accretion disks \cite{Li02,WXL02}, and magnetohydrodynamic flows in
geometrically thick disks \cite{Ca86,TNTT90}. Accretion disk
instabilities were also recently discussed in Ref. \cite{JC}. The black hole
spin evolution due to accretion, in connection with radiation efficiency,
when both jets and magnetic fields are present was investigated in Ref. \cite%
{KGB}. Another approach for studying magnetosphere models of Schwarzschild
black holes with non-magnetized accretion disks consists in solving the
Grad-Shafranov equation, derived from the Einstein-Maxwell equations. A
stationary axisymmetric force-free magnetosphere in a Schwarzschild geometry
was studied in Ref. \cite{Uzdensky}. In this model the black hole is
connected by the magnetic field lines to a thin Keplerian disk. A uniform
magnetic field at the event horizon was found to be a reasonable assumption
in the nonrotating limit. It also turned out that a uniform radial magnetic
field is still an excellent approximation for slowly rotating Keplerian
disks.

A static and spherically symmetric black hole immersed in an asymptotically
uniform magnetic field was presented by Preston and Poisson \cite{PP}. An
accretion disk in this geometry will have slightly modified properties
compared to the vacuum case, due to the weak magnetic field of this
space-time. This is what we propose to study in this paper.

The Preston-Poisson metric was derived based on the light-cone gauge
introduced in Ref.~\cite{PP} for perturbed Schwarzschild black holes. This
gauge preserves three convenient properties of the Eddington-Finkelstein
coordinates of the Schwarzschild metric. Namely (i) the advanced-time
coordinate $v$ is constant on incoming light cones that converge toward the
center, (ii) the polar and azimuthal angles are constant on the null
generators of each light cone, (iii) the radial coordinate is an
affine-parameter distance along each generator. In the unperturbed scenario
there is a fourth property, (iv) the radial coordinate $r$ is an areal
radius or curvature coordinate \cite{KucharSchw}, defined by the condition
that the area of the 2-spheres with constant $r$ is $4\pi r^{2}$ as in flat
space. This fourth property is however not obeyed in a generic perturbed
scenario.

As an application of the formalism Preston and Poisson have derived the
perturbations of the Schwarzschild metric arising from the immersion of the
black hole into an asymptotically uniform magnetic field. By carefully
performing the integration, in top of the modifications induced by the
magnetic field, they derive an additional tidal perturbation, due to distant
structures. Thus the Preston-Poisson perturbative solution represents a
magnetized black-hole space-time in which the tidal gravity is not directly
tied to the magnetic field. In this sense it is a generalization of the
exact, two-parameter Schwarzschild-Melvin metric family, where all
perturbations are of magnetic origin \cite{SchMelvin}.

The magnetic field generates a quadrupolar deformation of the event horizon.
Despite the penetration of the magnetic field lines below the horizon its
area stays unchanged. This is a combined consequence of the Hawking-Hartle
formula \cite{HH}, according to which the change of the area during the
quasi-static perturbation is determined by the flux of energy $T^{rr}$
crossing the horizon; and of the particular form of $T^{rr}$ for this
specific black hole, which vanishes on the horizon (at least to $B^{2}$
order, where $B$ is the strength of the magnetic field).

In Ref. \cite{Konoplya} Konoplya has rewritten the Preston-Poisson metric
into a diagonal form by a suitable redefinition of the radial variable and a
replacement of the null coordinate by a temporal variable. For the latter a
tortoise-type transformation was employed. Then he has analyzed the motion
of particles around such black holes. He has studied equatorial orbits and
found that the tidal perturbations from surrounding sources have significant
influence on the motion of test particles. The time delay and the bending
angle characterizing massless particles together with the binding energy of
massive particles have increased, while the radius of the innermost stable
circular orbit is decreased due to the presence of tidal forces.

Our aim here is to study the accretion processes onto Preston-Poisson black
holes, which incorporate both magnetic and tidal perturbations of the
Schwarzschild black hole. In Section 2 we present a short summary of the
accretion process in the absence of the magnetic fields.

In Section 3 we briefly review the Preston-Poisson metric both in its
original light-cone gauge form, and in the coordinates presented in Ref.
\cite{Konoplya}, employing also the analysis of the curvature invariants
from Appendix \ref{curvature}. We establish the radial range over which this
geometry describes a perturbed Schwarzschild black hole.

We analyze the geodesic motion in the equatorial plane in terms of an
effective potential in Section 4.

Here we also present the numerical study of the modifications induced by the
magnetic field and tidal perturbations in the disk radiation, temperature
profile, spectrum, luminosity and energy conversion efficiency for the
Preston-Poisson black hole. For this we employ the explicit form of the
energy-momentum tensor given in Appendix \ref{Tab}.

Finally, Section 5 contains the Concluding Remarks.

\section{The accretion process}

In the steady state accretion disk model physical quantities describing
matter fields are averaged over the characteristic time scale $\Delta t$,
total azimuthal angle $2\pi $ and accretion disk height $H$ (defined by its
maximum half thickness).

The matter in the accretion disk is modeled by an anisotropic fluid, where
the density $\rho _{0}$ of the rest mass (the specific heat is neglected),
the energy flow vector $q^{a}$ and stress tensor $t^{ab}$ are defined in the
averaged rest-frame of the orbiting plasma with 4-velocity $u^{a}$. The
invariant algebraic decomposition of the stress-energy tensor is
\begin{equation*}
T^{ab}=\rho _{0}u^{a}u^{b}+2u^{(a}q^{b)}+t^{ab}\;,
\end{equation*}%
where $u^{a}q_{a}=0=u^{a}t_{ab}$.

In this hydrodynamic approximation Page and Thorne \cite{PaTh74} have
derived the law of rest mass conservation, stating that the time averaged
rate of rest mass accretion is independent of the radius: $\dot{M_{0}}\equiv
dM_{0}/dt=-2\pi r\Sigma u^{r}=\mbox{const}$. (Here $t$ and $r$ are the time
and radial coordinates and $\Sigma $ is the averaged surface density). The
integral form of the conservation laws of angular momentum and energy was
also derived by averaging the continuity equation and the total divergence
of the density-flux 4-vectors $J^{a}=T^{ab}\varphi _{b}$ (angular momentum
density-flux) and $E^{a}=-T^{ab}t_{b}$ (energy density-flux), respectively.
Here $\varphi ^{a}=\partial /\partial \varphi $ and $t^{a}=\partial
/\partial t$ are the Killing vectors of the axially symmetric geometry and $%
\varphi $ is the azimuthal coordinate.

From the integral form of the conservation laws of energy and
angular-momentum and the energy-angular momentum relation $\widetilde{E}%
_{,r}=\Omega \widetilde{L}_{,r}$, Page and Thorne have expressed the
time-averaged vertical component $F$ (the photon flux) of the energy flow
vector $q^{a}$ as
\begin{equation}
F(r)=\frac{\dot{M}_{0}}{4\pi \sqrt{-g}}\frac{-\Omega _{,r}}{(\widetilde{E}%
-\Omega \widetilde{L})^{2}}\int_{r_{ms}}^{r}(\widetilde{E}-\Omega \widetilde{%
L})\widetilde{L}_{,r}dr\;.  \label{F}
\end{equation}%
Here $\widetilde{E}$, $\widetilde{L}$ and $\Omega =d\varphi /dt$ are the
specific energy, specific angular-momentum and angular velocity of the
orbiting plasma particles with respect to the coordinate time $t$. The above
formula is valid under the assumption that the torque of the infalling
matter on the disk vanishes at the inner edge of the disk (since the
accreting matter reaching the marginally stable orbit $r_{ms}$ falls freely
into the hole and cannot exert any considerable torque).

Supposing that the electron-scattering opacity is negligible and the
accretion disk is \textit{optically} thick, the disk surface radiates a
black body spectrum. Then the surface temperature $T(r)$ of the disk is
given by $F(r)=\sigma T^{4}(r)$, with the Stefan-Boltzmann constant $\sigma $%
. The disk luminosity $\mathcal{L}\left( \omega \right) $ is calculated as
function of $T$ (which is in turn expressed in terms of the thermal photon
flux) as
\begin{equation}
\mathcal{L}\left( \omega \right) =\frac{4\omega ^{3}}{\pi }\cos \iota
\int_{r_{ms}}^{\infty }\frac{rdr}{\exp \left( \omega /T\right) -1},
\label{L}
\end{equation}%
where $\iota $ is the inclination angle of the disk with respect to the line
of sight. For simplicity we assume $\cos \iota =1$.

Another important characteristics of the mass accretion process is the
efficiency with which the central object converts rest mass into outgoing
radiation. The efficiency is defined as the ratio of two rates evaluated at
infinity: the rate of the radiated energy of photons escaping from the disk
surface to infinity over the rate at which mass-energy is transported to the
black hole \cite{NoTh73,PaTh74}. If all emitted photons escape to infinity,
the efficiency is given in terms of the specific energy measured at the
marginally stable orbit $r_{ms}$ as
\begin{equation}
\epsilon =1-\widetilde{E}_{ms}.  \label{epsilon}
\end{equation}

For Schwarzschild black holes the efficiency $\epsilon $ is about $6\%$,
irrespective of whether photon capture by the black hole is taken into
account or not. However, for rapidly rotating black holes, the efficiency $%
\epsilon $ is found to be $42.3\%$, decreasing slightly to $40\%$ with
photon capture by the black hole included \cite{Th74}.

\section{Perturbed Schwarzschild black hole region of the Preston-Poisson
space-time}

In this section we review the Preston-Poisson metric, both in the original
light-cone gauge coordinates employed in \cite{PP} and in the coordinates
introduced in Ref. \cite{Konoplya}. The latter is essential in studying the
accretion processes in the remaining part of the paper. Then we analyze the
equatorial geometry and we establish the radial range where the
interpretation of a perturbed Schwarzschild black hole holds.

The Preston-Poisson metric represents a perturbed Schwarzschild black hole
with perturbations caused by (i) an asymptotically uniform magnetic field $B$
and (ii) independent tidal effects, described by a parameter $K$. The
perturbations are such that the perturbed space-time is stationary and
axially symmetric. In lowest order the rotational Killing vector of the
space-time can be used to define the asymptotically uniform magnetic field
\cite{Wald}, through the 4-potential \cite{PP}%
\begin{equation}
A^{a}=\left( 0,0,0,B/2\right) \ .  \label{Aa}
\end{equation}

The metric given in the light-cone gauge {[}in the Eddington-Finkelstein
type coordinates $(v,r,\theta ,\phi )$] is%
\begin{eqnarray}
g_{vv} &=&-f-\tfrac{1}{9}B^{2}r(3r-8M)-\left[ \tfrac{1}{9}%
B^{2}(3r^{2}-14Mr+18M^{2})-K(r-2M)^{2}\right] (3\cos ^{2}\theta -1)\ ,
\notag \\
g_{vr} &=&1\ ,  \notag \\
g_{v\theta } &=&\left[ \tfrac{2}{3}B^{2}(r-3M)-2K(r-2M)\right] r^{2}\sin
\theta \cos \theta \ ,  \notag \\
g_{\theta \theta } &=&r^{2}-\left[ \tfrac{1}{3}%
B^{2}r^{2}-B^{2}M^{2}-K(r^{2}-2M^{2})\right] r^{2}\sin ^{2}\theta \ ,  \notag
\\
g_{\varphi \varphi } &=&r^{2}\sin ^{2}\theta -\left[ \tfrac{1}{3}%
B^{2}r^{2}+B^{2}M^{2}+K(r^{2}-2M^{2})\right] r^{2}\sin ^{4}\theta ~,
\label{PPEdFink}
\end{eqnarray}%
where $f=1-{2M}/{r}$ and $M$ is the mass of the corresponding Schwarzschild
black hole. This form of the metric is accurate up to $(B^{2},K)$ order. {[}%
These are Eqs. (3.43)-(3.47) of Ref. \cite{PP} with the change of notation $%
\mathcal{E}\rightarrow K$. They are also given as Eqs.(3)-(6) of Ref. \cite%
{Konoplya}, however the last term of the respective Eq.(3) should be
corrected as $-K(r-2M)^{2}$, while the last term of Eq.(4) as $-2K(r-2M)$.]

The area of spheres with radius $r$ is modified by the magnetic field as
\begin{equation}
A_{r=\text{const}}=2\pi\int_{0}^{\pi}\sqrt{g_{\theta\theta}g_{\varphi%
\varphi}-g_{\theta\varphi}^{2}}d\theta=4\pi r^{2}\left(1-\tfrac{2}{9}%
B^{2}r^{2}\right)~,
\end{equation}
thus $r$ fails to be a curvature coordinate.

\begin{figure}[tbp]
\includegraphics[width=7cm]{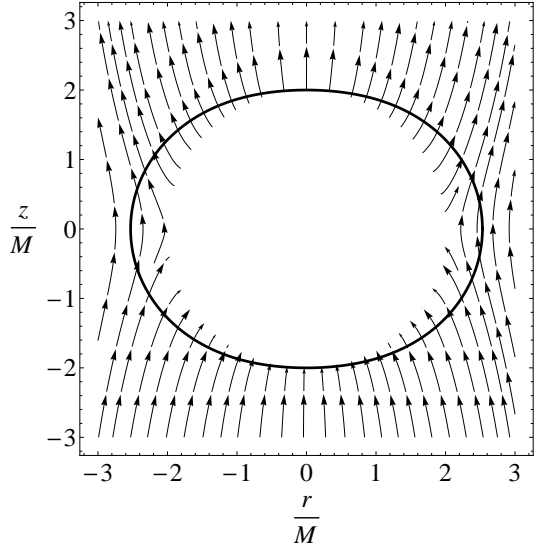}
\caption{The quadrupolar deformation of the horizon and the
structure of the magnetic field. (For illustrational purposes a
large value of the magnetic field $B=10^{-0.2}M^{-1}$ was chosen,
yielding $\protect\eta _{B}=1.6$, which is outside the perturbative
regime.)} \label{MagneticField1}
\end{figure}

In the perturbed space-time $\partial /\partial t=(1,0,0,0)$ remains a
Killing vector. Due to Hawking's strong rigidity theorem the event horizon
is given by the condition that $\partial /\partial t$ becomes null on it,
i.e. $\partial /\partial t\cdot \partial /\partial t\equiv g_{\mathbf{tt}}=0$%
. Under the magnetic perturbation the event horizon acquires a quadrupolar
deformation:
\begin{equation}
r_{H}\left( \theta \right) =2M\left( 1+\tfrac{2}{3}M^{2}B^{2}\sin ^{2}\theta
\right) ~,  \label{rH}
\end{equation}%
but quite remarkably its area is unchanged (to linear order in the
perturbations) as compared to the Schwarzschild black hole:%
\begin{equation}
A_{H}=16\pi M^{2}\ .  \label{AH}
\end{equation}%
The quadrupolar deformation of the horizon and the magnetic field topology
are illustrated on Fig \ref{MagneticField1}.

By introducing a temporal variable with a tortoise-like transformation and
passing to a new radial coordinate $\bar{r}$ as%
\begin{eqnarray}
v &=&t+\bar{r}+2M\ln \left\vert \frac{\bar{r}}{2M}-1\right\vert \ ,
\label{t} \\
r &=&\bar{r}\{1+\tfrac{1}{3}B^{2}\bar{r}[M+(\bar{r}-3M)\cos ^{2}\theta ]-%
\tfrac{1}{3}K\bar{r}(\bar{r}-2M)(3\cos ^{2}\theta -1)\}\ ,  \label{rbar}
\end{eqnarray}%
Konoplya has rewritten the Preston-Poisson metric in a diagonal form \cite%
{Konoplya},
\begin{eqnarray}
g_{tt} &=&-\left( 1-\frac{2M}{\bar{r}}\right) +\tfrac{1}{3}K(4M-3\bar{r})(2M-%
\bar{r})(3\cos ^{2}\theta -1)-\tfrac{1}{3}B^{2}[3(2M-\bar{r})\cos ^{2}\theta
-2M](2M-\bar{r})\ ,  \notag \\
g_{\bar{r}\bar{r}} &=&\left( 1-\frac{2M}{\bar{r}}\right) ^{-1}-K\frac{\bar{r}%
^{2}(4M-3\bar{r})(3\cos ^{2}\theta -1)}{3(2M-\bar{r})}+B^{2}\frac{\bar{r}%
^{2}[3(2M-\bar{r})\cos ^{2}\theta -2M]}{3(2M-\bar{r})}\ ,  \notag \\
g_{\theta \theta } &=&\bar{r}^{2}+\tfrac{1}{3}K\bar{r}^{2}[2(3\cos \theta
^{2}-1)(2M-\bar{r})\bar{r}-3(2M^{2}-\bar{r}^{2})\sin \theta ^{2}]  \notag \\
&&\quad -\tfrac{1}{3}B^{2}\bar{r}^{2}\{2[(3M-\bar{r})\cos \theta ^{2}-M]\bar{%
r}-(3M^{2}-\bar{r}^{2})\sin \theta ^{2}\}\ ,  \notag \\
g_{\varphi \varphi } &=&\bar{r}^{2}\sin \theta ^{2}+\tfrac{1}{3}K\bar{r}%
^{2}[2(3\cos \theta ^{2}-1)(2M-\bar{r})\bar{r}+3(2M^{2}-\bar{r}^{2})\sin
\theta ^{2}]\sin \theta ^{2}  \notag \\
&&\quad -\tfrac{1}{3}B^{2}\bar{r}^{2}\{2[(3M-\bar{r})\cos \theta ^{2}-M]\bar{%
r}+(3M^{2}+\bar{r}^{2})\sin \theta ^{2}\}\sin \theta ^{2}\ .
\label{eq:Konoplya metric}
\end{eqnarray}%
The diagonal form of the metric allows for a simpler description of the
motion of particles and the accretion process. We find however that these
new coordinates do not preserve all the convenient properties of the
Eddington-Finkelstein coordinates, namely, the radial coordinate $\bar{r}$
fails to coincide with an affine-parameter distance along the generators of
incoming light cones.

The transformation to the coordinates $\left\{ t,\bar{r},\theta,\phi\right\}
$ is $\theta$-dependent; for the equatorial plane it simplifies to
\begin{equation}
r_{\theta=\pi/2}=\bar{r}\left[1+\tfrac{1}{3}B^{2}\bar{r}M+\tfrac{1}{3}K\bar{r%
}(\bar{r}-2M)\right]~,
\end{equation}
while the metric in the equatorial plane reduces to
\begin{eqnarray}
g_{tt} & = & -\left(1-\frac{2M}{\bar{r}}\right)+\tfrac{1}{3}K(-8M^{2}+10M%
\bar{r}-3\bar{r}^{2})+\tfrac{2}{3}B^{2}M(2M-\bar{r})\ ,  \notag \\
g_{\bar{r}\bar{r}} & = & \left(1-\frac{2M}{\bar{r}}\right)^{-1}+K\frac{\bar{r%
}^{2}(4M-3\bar{r})}{3(2M-\bar{r})}-B^{2}\frac{2M\bar{r}^{2}}{3(2M-\bar{r})}\
,  \notag \\
g_{\theta\theta} & = & \bar{r}^{2}+\tfrac{1}{3}K\bar{r}^{2}(-6M^{2}-4M\bar{r}%
+5\bar{r}^{2})+\tfrac{1}{3}B^{2}\bar{r}^{2}(3M^{2}+2M\bar{r}-\bar{r}^{2})\ ,
\notag \\
g_{\varphi\varphi} & = & \bar{r}^{2}+\tfrac{1}{3}K\bar{r}^{2}(6M^{2}-4M\bar{r%
}-\bar{r}^{2})+\tfrac{1}{3}B^{2}\bar{r}^{2}(-3M^{2}+2M\bar{r}-\bar{r}^{2})\ .
\label{geq}
\end{eqnarray}
These are Eqs. (10)-(13) of Konoplya, however, the last term in the first
line of Eq. (13) of \cite{Konoplya} is corrected as $+\bar{r}^{2}$.

We find, remarkably, that the coordinate $\bar{r}$ is a curvature
coordinate,
\begin{equation}
A_{\bar{r}=\text{const}}=4\pi \bar{r}^{2}\ .
\end{equation}%
With the above form of $g_{tt}$ the horizon is described by its unperturbed
value
\begin{equation}
\bar{r}_{H}=2M\ ,  \label{eq:new horizon}
\end{equation}%
a result we have checked either by direct computation, or by inserting the
expression of the event horizon (\ref{rH}) into the inverse of the
coordinate transformation (\ref{rbar}). The area of the event horizon
computed in these coordinates confirms Eq.~(\ref{AH}).

Nevertheless, the metric at the horizon is in fact perturbed, as can be seen
by an explicit computation of the curvature invariants, given in Appendix %
\ref{curvature}. As the Kretschmann scalar $R_{abcd}R^{abcd}$ and the Euler
scalar ${}^{\ast }R_{\ abcd}^{\ast }R^{abcd}$ show an explicit $\theta $%
-dependence, we conclude that the horizon acquires the quadrupolar
deformation.

A glance at Eq.~(\ref{eq:Konoplya metric}) shows that the interpretation of
the Preston-Poisson metric as a perturbed black hole withstands only while
the parameters $\eta _{B}=B^{2}\bar{r}^{2}$ and $\eta _{K}=K\bar{r}^{2}$
stay small. Thus we may interpret the metric~(\ref{eq:Konoplya metric}) as a
perturbed Schwarzschild black hole only for $\bar{r}$ in the range%
\begin{equation}
2M\lesssim \bar{r}\lesssim r_{1}~,  \label{range}
\end{equation}%
with%
\begin{equation}
r_{1}\approx \min \left( \eta _{K}^{1/2}K^{-1/2},\eta
_{B}^{1/2}B^{-1}\right) \ll \min (K^{-1/2},B^{-1})~.  \label{r1}
\end{equation}%
\textbf{(}The condition that the parameters $\eta _{B}$ and $\eta _{K}$
should stay small in order the perturbative treatment to hold will determine
for any pair $B,\ K$ the value of $r_{1}$.)

In the range (\ref{range}) the metric perturbations of the black hole due to
the tidal force and the magnetic field remain small.

In the study of thin accretion disks it is convenient to introduce the
coordinate $z=\bar{r}\cos \theta \approx \bar{r}\left( \theta -\pi /2\right)
$ instead of the polar angle $\theta $. Therefore, the geometry describing
the space-time region where the disk is located is characterized by the
metric components $g_{tt}$, $g_{\bar{r}\bar{r}}$, $g_{\varphi \varphi }$ and
\begin{equation}
g_{zz}=\frac{g_{\theta \theta }}{\bar{r}^{2}}\ ,  \label{geqz}
\end{equation}%
given in Eq. (\ref{geq}) to zeroth order in $z$. We note that the back
reaction of the disk on the static Preston-Poisson geometry is neglected.

\section{Modifications in the accretion induced by the magnetic field and
tidal parameter}

To simplify our notation from now on we suppress the overbar from the
Konoplya radial variable.

\subsection{Orbital motion in the equatorial plane}

Here we analyze the radial dependence of the angular velocity $\Omega $,
specific energy $\widetilde{E}$ and specific angular momentum $\widetilde{L}$
of particles moving in circular and equatorial orbits. The axially symmetric
geometry is described by the metric~(\ref{geq}) and (\ref{geqz}). In this
approximation the off-diagonal components of the metric vanish and the
geodesic equations for particles orbiting in the equatorial plane of the
black hole can be written as
\begin{eqnarray}
g_{tt}^{2}\left( \frac{dt}{d\lambda }\right) ^{2} &=&\widetilde{E}^{2}\;,
\notag \\
g_{\varphi \varphi }^{2}\left( \frac{d\varphi }{d\lambda }\right) ^{2} &=&%
\widetilde{L}^{2}\;,  \notag \\
(g_{tt}g_{rr})^{2}\left( \frac{dr}{d\lambda }\right) ^{2}+V_{eff}^{2}(r) &=&%
\widetilde{E}^{2}\;,
\end{eqnarray}%
where $\lambda $ is the affine parameter, and the effective potential $%
V_{eff}(r)$ is given by
\begin{equation}
V_{eff}^{2}(r)\equiv g_{tt}\left( 1+\frac{\widetilde{L}^{2}}{r^{2}}\right)
\;.  \label{V2}
\end{equation}

From the conditions $V_{eff}=\widetilde{E}^{2}$ and $V_{eff~,r}=0$, which
define the circular orbits around the central object we obtain\footnote{%
From the normalization $u^{a}u_{a}=-1$ we get $u^{t}=(-g_{tt}-\Omega
^{2}g_{\varphi \varphi })^{-1/2}$, which can be inserted into the
expressions $\widetilde{E}=-u_{t}=g_{tt}u^{t}$ and $\widetilde{L}=-u_{\phi
}=g_{\varphi \varphi }u^{\phi }=\Omega g_{t\varphi }u^{t}$. These give Eqs. (%
\ref{E}) and (\ref{Ltilde}).}
\begin{eqnarray}
\Omega &=&\frac{d\varphi }{dt}=\frac{u^{\varphi }}{u^{t}}=\sqrt{\frac{%
-g_{tt,r}}{g_{\varphi \varphi ,r}}}\;,  \label{Omega} \\
\widetilde{E} &=&-u_{t}=-\frac{g_{tt}}{\sqrt{-g_{tt}-g_{\varphi \varphi
}\Omega ^{2}}}\;,  \label{E} \\
\widetilde{L} &=&u_{\varphi }=\frac{g_{\varphi \varphi }\Omega }{\sqrt{%
-g_{tt}-g_{\varphi \varphi }\Omega ^{2}}}  \label{Ltilde}
\end{eqnarray}

Substituting Eq. (\ref{Ltilde}) into $V_{eff}$ we obtain $V_{eff}=\widetilde{%
E}^{2}$. Inserting Eq. (\ref{Ltilde}) into $V_{eff,r}=0$ the explicit
expression (\ref{Omega}) for the angular velocity is recovered. The
condition $V_{eff~,rr}=0$ gives the marginally stable orbit (innermost
stable circular orbit) $r_{ms}$.

\begin{figure}[tbp]
\includegraphics[width=8.7cm]{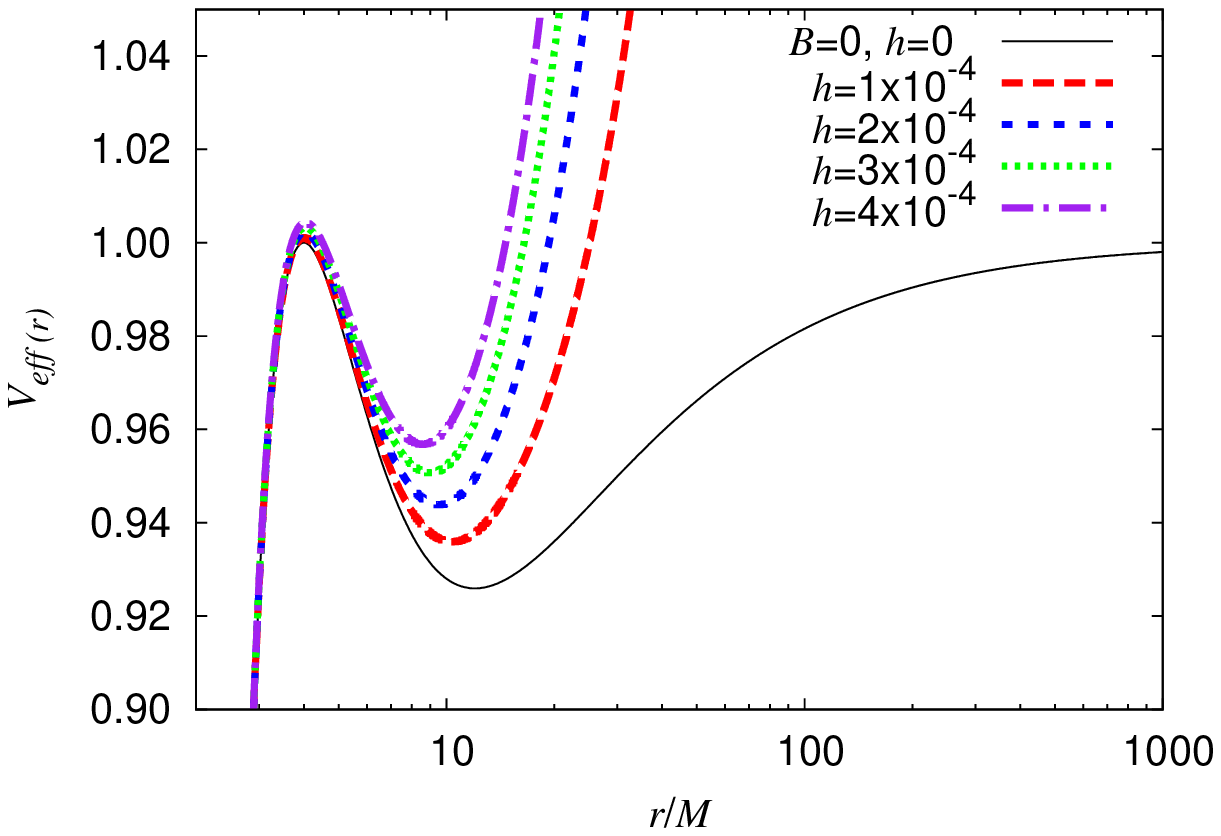} %
\includegraphics[width=8.7cm]{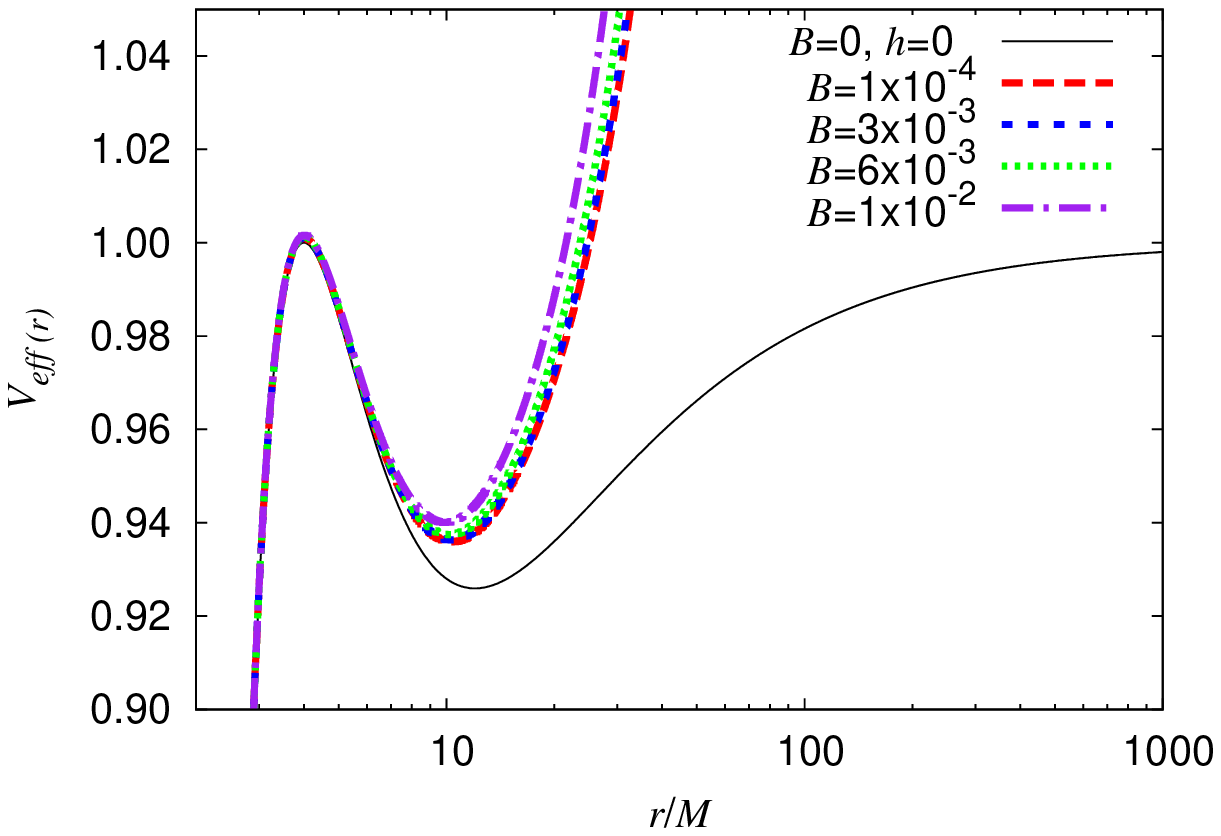}
\caption{The effective potential for a perturbed Scwarzschild black hole of
a total mass $M$ and specific angular momentum $\widetilde{L}=4M$. The solid
line is the effective potential for a Schwarzschild black hole with the same
total mass ($B=0$ and $K=0$). On the left plot $B$ is set to $10^{-4}M^{-1}$
and the parameter $h$ is running, while on the right $h=10^{-4}M^{-2}$ is
fixed and different values of $B$ are taken.}
\label{V_eff_Fig}
\end{figure}

As a first step we consider the radial dependence of the effective potential
(\ref{V2}) of the perturbed Schwarzschild black hole and compare it with the
non-perturbed case. In the left plot of Fig \ref{V_eff_Fig} we present the
radial profile of the potential with different values of the tidal parameter
$K$ in a magnetic field with fixed field strength of $B=10^{-4}M^{-1}$. The
parameter $K$ is given as $K=B^{2}/2+h$, with $h$ running between $%
10^{-4}M^{-2}$ and $4\times 10^{-4}M^{-2}$. Due to the presence of the
asymptotically uniform magnetic field the perturbed Schwarzschild potential
fails to be asymptotically flat, it actually diverges for $r\rightarrow
\infty $.

Increasing the parameter $K$ (or $h$) we also increase the steepness with
which the potential tends to spatial infinity as we are receding from the
central object. We have also checked that the divergent behavior of the
potential appears also if only one of the perturbations is present.

On the right plot of Fig \ref{V_eff_Fig} we have fixed $K$ and set the
magnetic field strength $B$ to $10^{-4}M^{-1}$, $3\times10^{-3}M^{-1}$, $%
6\times10^{-3}M^{-1}$ and $10^{-2}M^{-1}$, respectively. The variation of $B$
modifies the steepness of how $V_{eff}$ diverges for $r\rightarrow\infty$.
With increasing field strength the effective potential diverges faster in
the spatial infinity.

\begin{figure}[tbp]
\includegraphics[width=8.7cm]{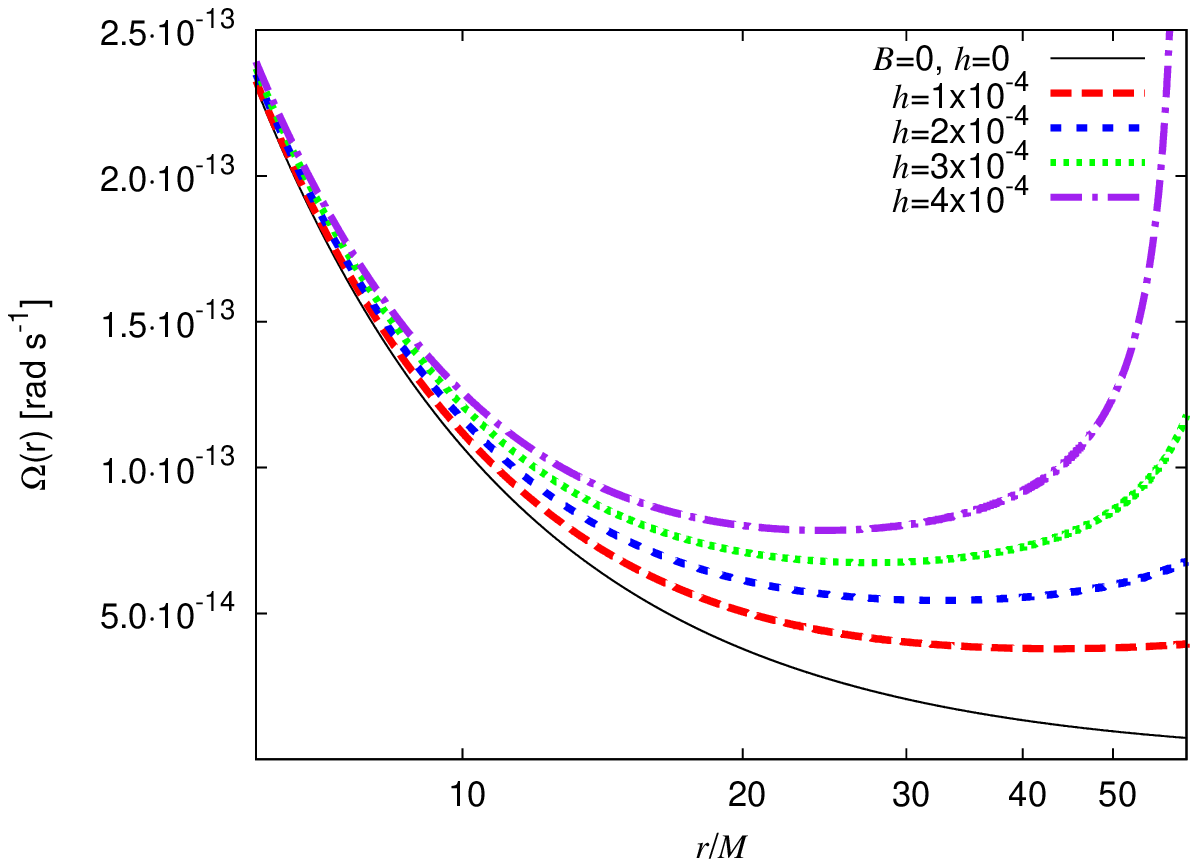} %
\includegraphics[width=8.7cm]{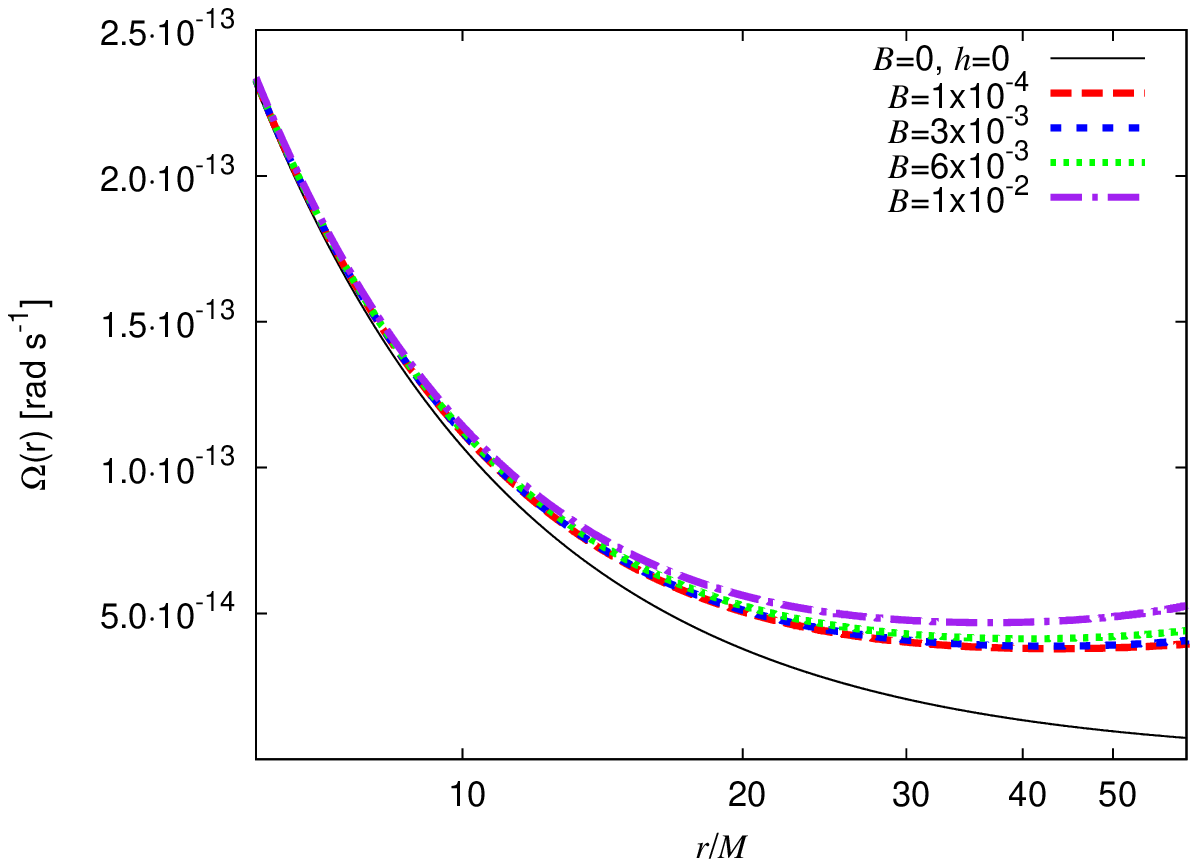}\newline
\includegraphics[width=8.7cm]{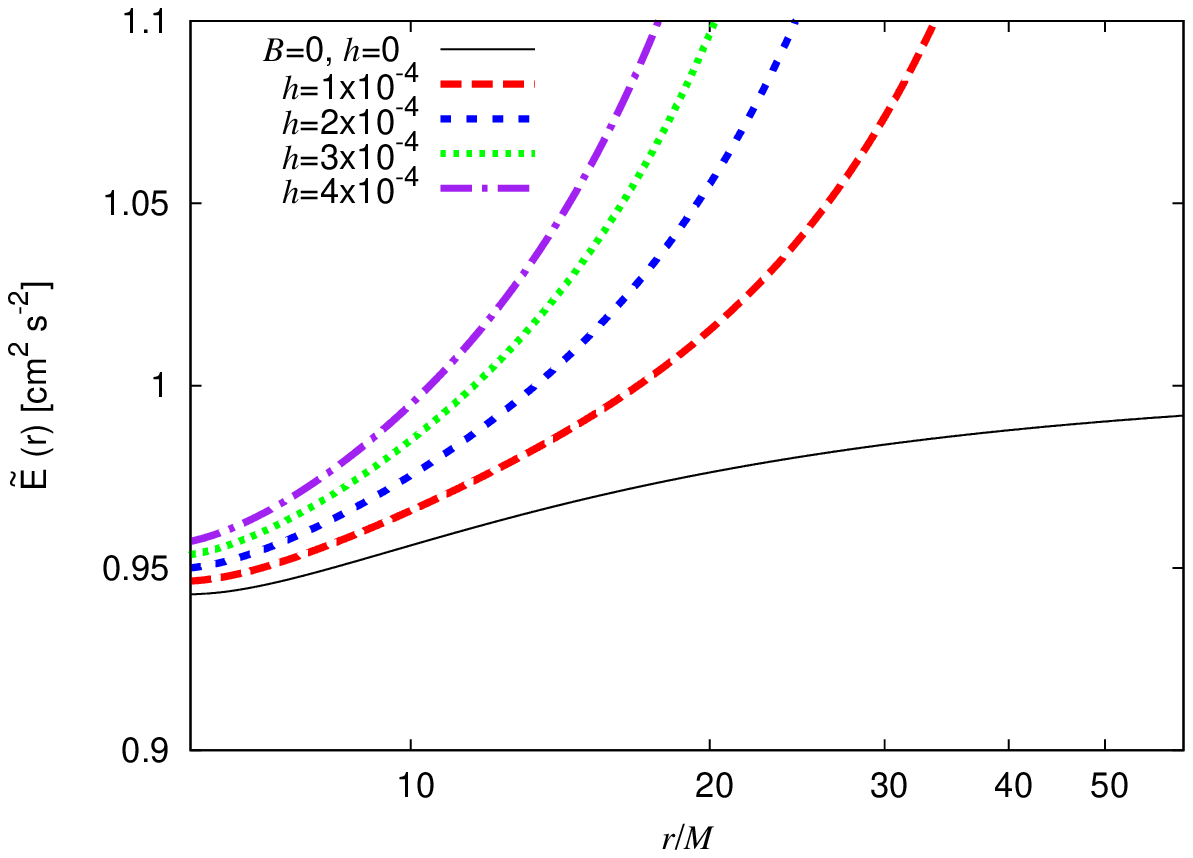} %
\includegraphics[width=8.7cm]{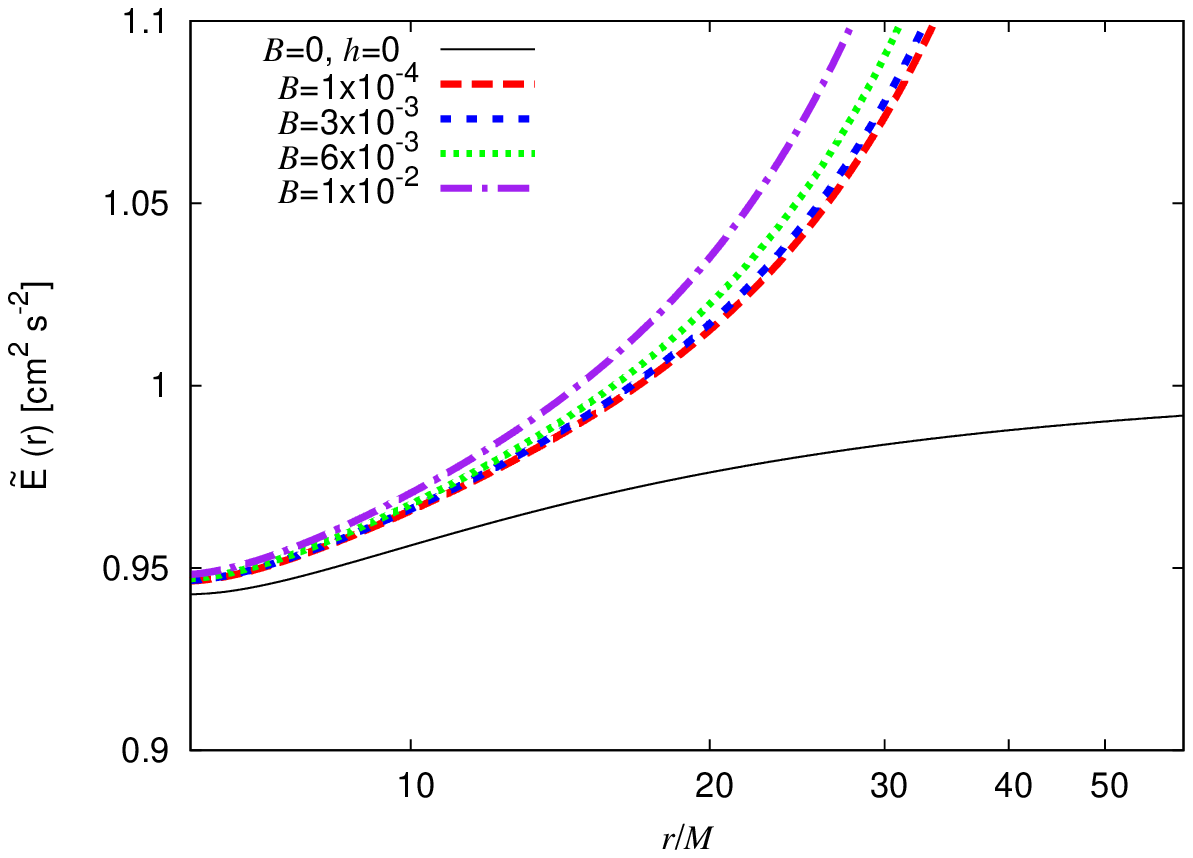}\newline
\includegraphics[width=8.7cm]{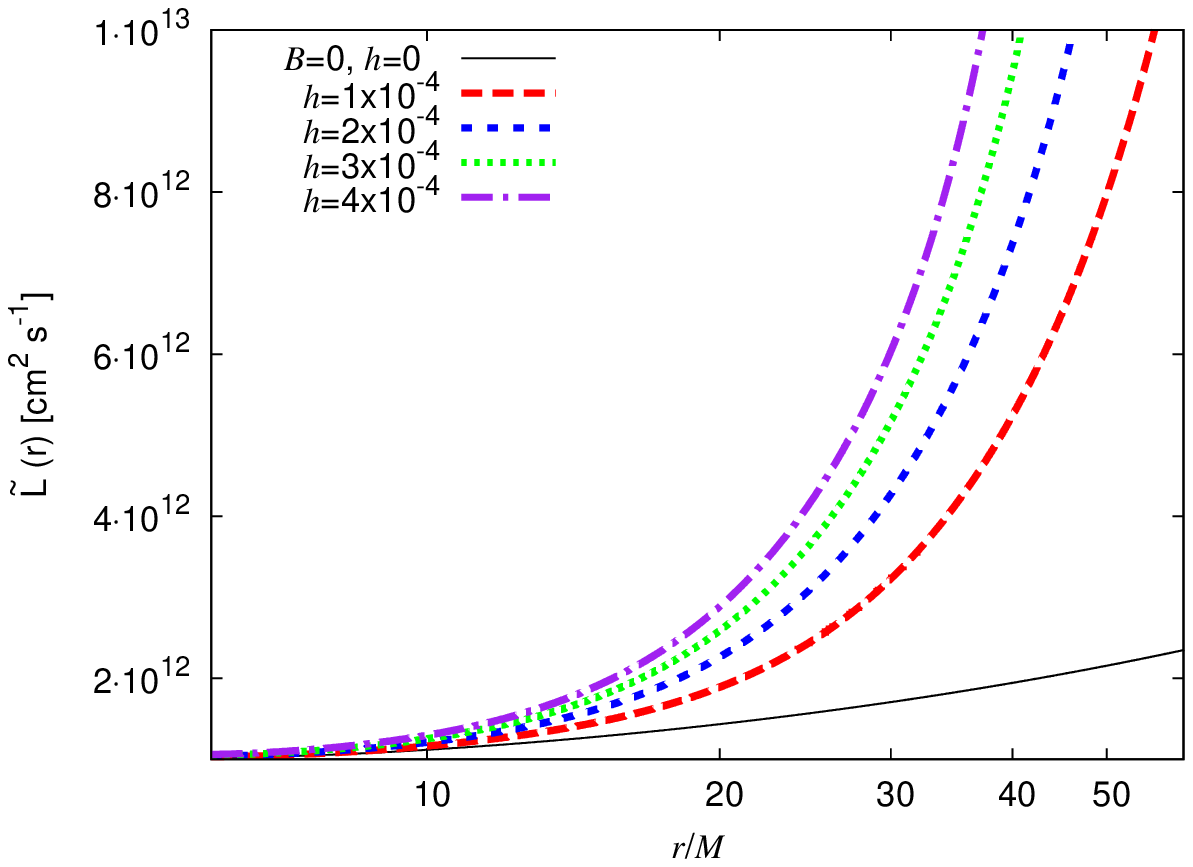} %
\includegraphics[width=8.7cm]{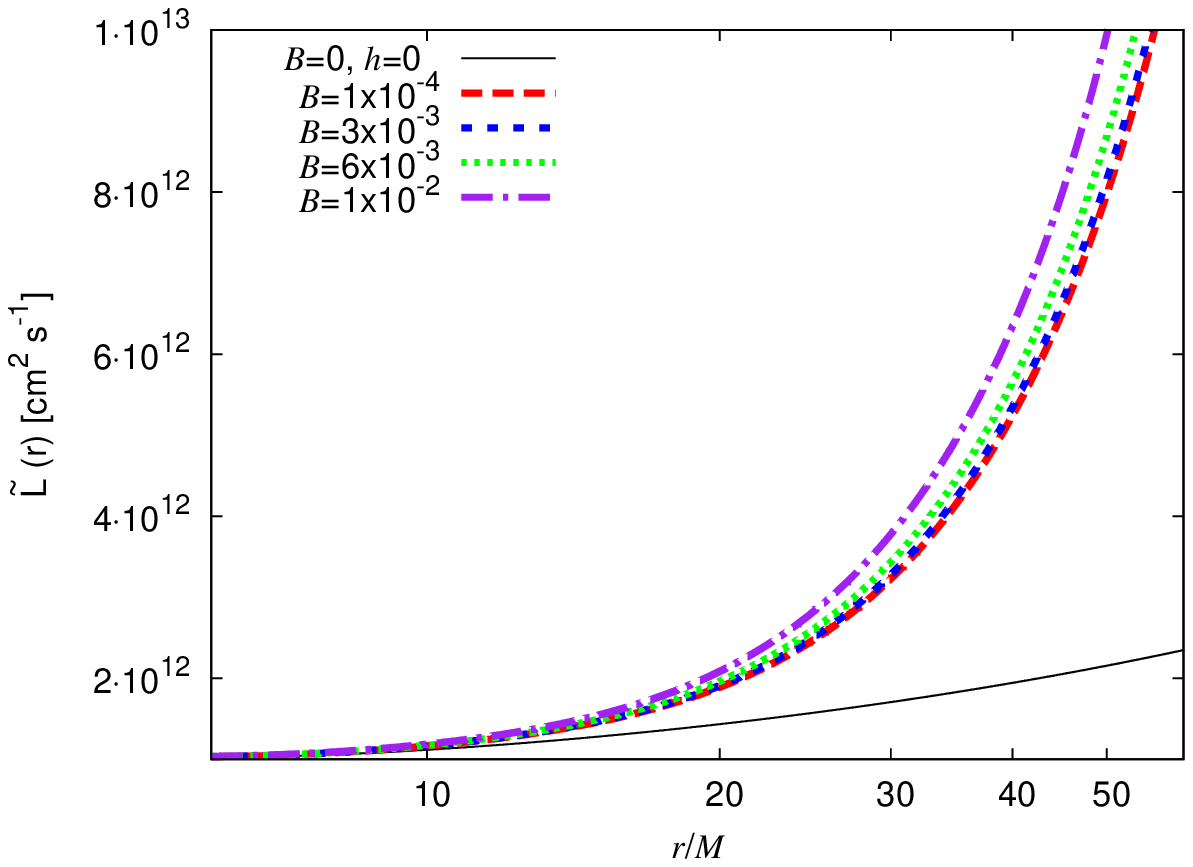}
\caption{The angular velocity $\Omega $, specific energy $\widetilde{E}$ and
specific angular momentum $\widetilde{L}$ of particles orbiting around a
perturbed Schwarzschild black hole of total mass $M$. The solid line is the
effective potential for a Schwarzschild black hole with the same total mass (%
$B=0$ and $K=0$). On the left plot $B$ is set to $10^{-4}M^{-1}$ and the
parameter $h$ is running, while on the right $h=10^{-4}M^{-2}$ is fixed and
different values of $B$ are taken.}
\label{OmEL_Fig}
\end{figure}

In Fig \ref{OmEL_Fig} we present the radial dependence of the angular
frequency, specific energy and specific angular momentum of the orbiting
particle. All of these radial profiles indicate the perturbative presence of
the asymptotically uniform magnetic field. Close to the black hole the
rotational velocity $\Omega $ resembles the unperturbed Schwarzschild value.
For higher radii, however, each radial profile of $\Omega $ has a less steep
fall-off compared to the one for a standard accretion disk in the
non-perturbed system. Moreover, at certain radii $\Omega $ is starting to
increase. This unphysical model feature is explained in the following
subsection. The radial profiles of $\widetilde{E}$ and $\widetilde{L}$ are
also unbounded as $r\rightarrow \infty $.

\subsection{Photon flux and disk temperature}

By inserting Eqs. (\ref{Omega})-(\ref{Ltilde}) into Eq. (\ref{F}) and
evaluating the integral we obtain the flux over the entire disk surface.
This enables us to derive the temperature profile and spectrum of the disk.
As shown in Appendix \ref{Tab}, the components $T_{t}^{r}$ $T_{t}^{z}$, $%
T_{\varphi }^{r}$ and $T_{\varphi }^{z}$ of the energy-momentum tensor for
the magnetic field vanish. Since only these quantities appear in the
integral form of the conservation laws of energy and angular momentum
specified for the steady-state equatorial approximation, the magnetic field
does not contribute to the photon flux radiated by the accretion disk at
all. Therefore, for a Schwarzschild black hole with magnetic perturbation we
can employ the same flux formula as for vacuum.

\begin{figure}[tbp]
\includegraphics[width=8.7cm]{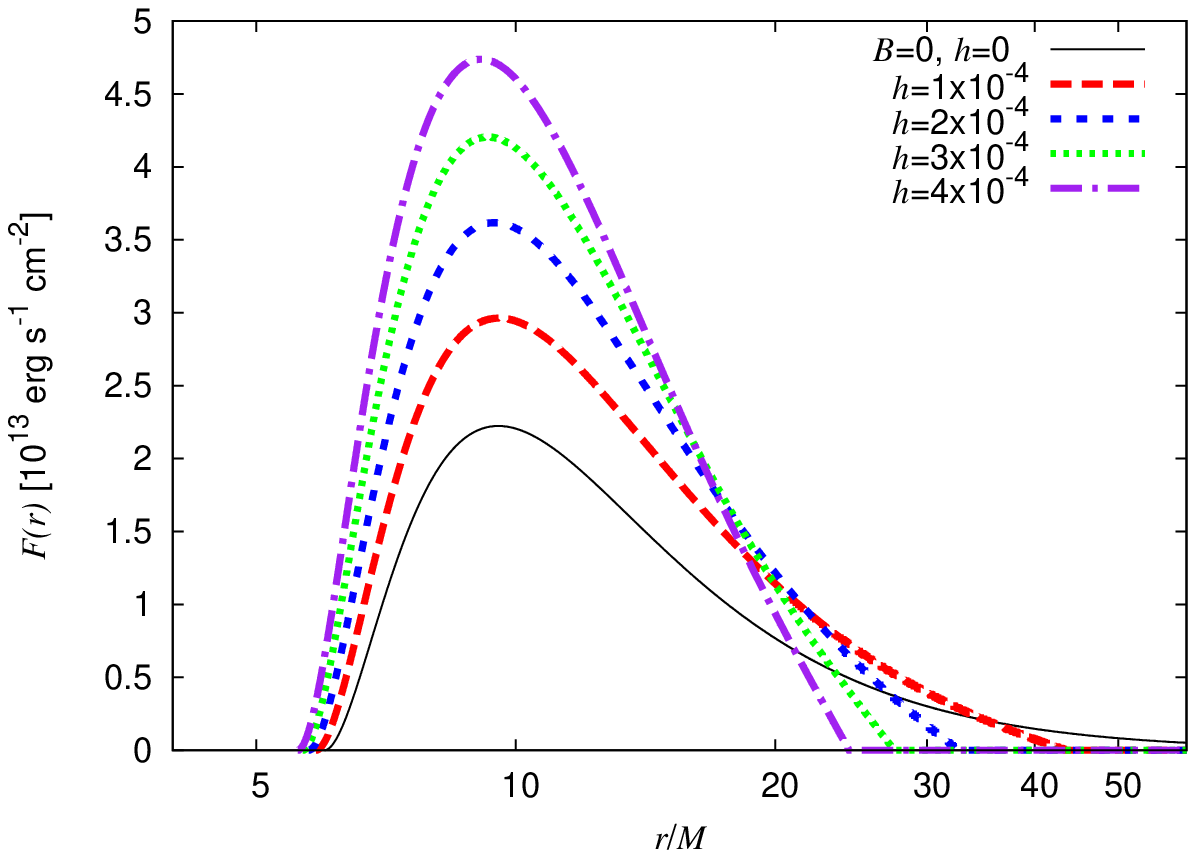} %
\includegraphics[width=8.7cm]{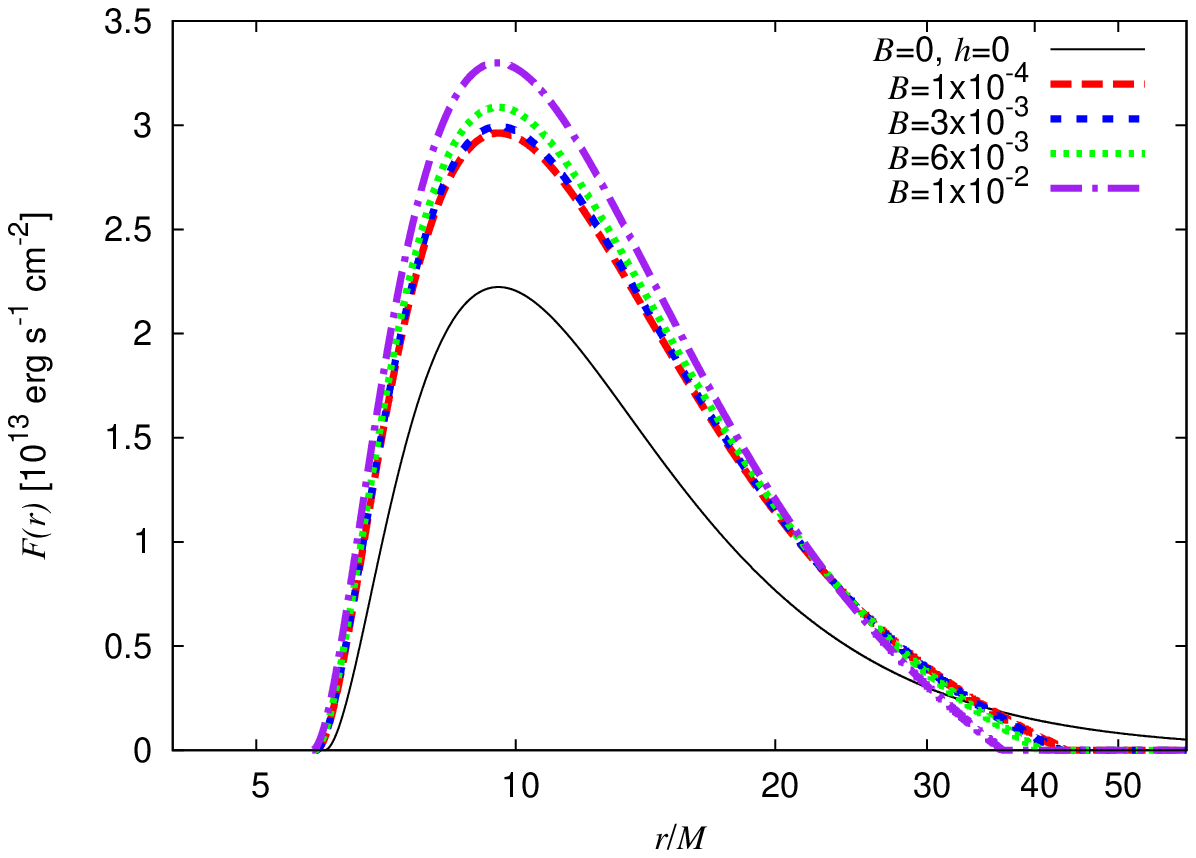}
\caption{The time-averaged flux radiated by the accretion disk around a
perturbed Schwarzschild black hole of total mass $M=2\times10^{6}M_{%
\bigodot} $. The accretion rate is $2.5\times10^{-6}M_{\bigodot}$/yr. The
solid line is the radiated flux for a Schwarzschild black hole with the same
total mass ($B=0$ and $K=0$). On the left plot $B$ is set to $10^{-4}M^{-1}$
and the parameter $h$ is running, while on the right $h=10^{-4}M^{-2}$ is
fixed and different values of $B$ are taken.}
\label{F_Fig}
\end{figure}

In Fig \ref{F_Fig} we plot the flux integral (\ref{F}) for a black hole
with mass $2\times 10^{6}M_{\odot }$ and an accretion rate of $2.5\times
10^{-6}M_{\odot }$/yr, with the same sets of values for the parameters $B$
and $K$ as for the effective potential. An increase of the parameters $B$
and $K$ results in smaller radii of both the marginally stable and largest
radius bound orbits, which shifts both the inner and outer edges of the
accretion disc towards the black hole. This can be seen in the plots of the
flux emitted by the disk where the radial flux profiles shift to lower
radii, compared with the radial distribution of $F(r)$ for an accretion disk
in Schwarzschild geometry.

A closer comparison of the flux profile shapes with $\Omega \left( r\right) $
on Fig \ref{OmEL_Fig} shows that for each parameter set the radius $r_{2}$
where $F\left( r\right) =0$ holds is precisely where $\Omega $ starts to
increase. Going further outwards the flux would turn negative, indicating
that the thin disk model breaks down at larger distances. Therefore we
should consider our thin accretion disk only extending between $r_{ms}$ and $%
r_{2}$, letting the condition $F\left( r\right) =0$ to determine the outer
radius of the thin disk.

On the graphs the approximate ranges of the perturbing parameters are $B\in
\left( 10^{-4},10^{-2}\right) M^{-1}$ and $K\in \left( 10^{-4},4\times
10^{-4}\right) M^{-2}$. For these parameters $\eta _{B}\in \left(
10^{-8},10^{-4}\right) \varepsilon ^{-2}$ and $\eta _{K}\in \left(
10^{-4},4\times 10^{-4}\right) \varepsilon ^{-2}$, where $\varepsilon =M/%
\bar{r}$ is the post-Newtonian parameter. Thus the maximum values of both
parameters $\eta _{B}$ and $\eta _{K}$ are of the order $10^{-4}\varepsilon
^{-2}$. As the accretion can be discussed only in the range where both the
magnetic field and tidal effects can be considered as perturbations of the
Schwarzschild black hole, both parameters $\eta _{B}$ and $\eta _{K}$ have
the upper limit $10^{-1}$. Therefore $10^{-4}\varepsilon ^{-2}\lesssim
10^{-1}$ and $\varepsilon =M/r\gtrsim 10^{-3/2}$. As a consequence, the
validity of the perturbed black hole picture holds in the range
\begin{equation}
2M\lesssim \bar{r}\lesssim 10^{3/2}M\approx 31M~.
\end{equation}%
The estimate of $r_{1}\approx 31M$ is in the range of the values for
$r_{2}$ readable from Fig \ref{F_Fig}.$\allowbreak $ We have seen
earlier that the perturbed black hole picture can be extended up to
$r_{1}$ only. Accretion disks are expected to exist only around
central objects. In the regions where the space-time is closer to a
uniform magnetic field perturbed by a black hole, rather than
vice-versa, it is to be expected that accretion disks should not
exist at all. As a first symptom of this, by increasing the radius
the thin disk approximation should become increasingly inaccurate.
This is the reason why the radius $r_{2}$, where the thin disk
approximation breaks down, has to be connected with $r_{1}$.

We note that $r_{2}$\ is more affected by the change of $B$ or $K$ then $%
r_{ms}$. For stronger perturbations the accretion disk is therefore located
closer to the black hole and its surface area is reduced.\textbf{\ }However,
the stronger magnetic field or higher value of $K$ increases the maximal
intensity of the radiation without causing any significant shift in the peak
of maximal flux.

\begin{figure}[tbp]
\includegraphics[width=8.7cm]{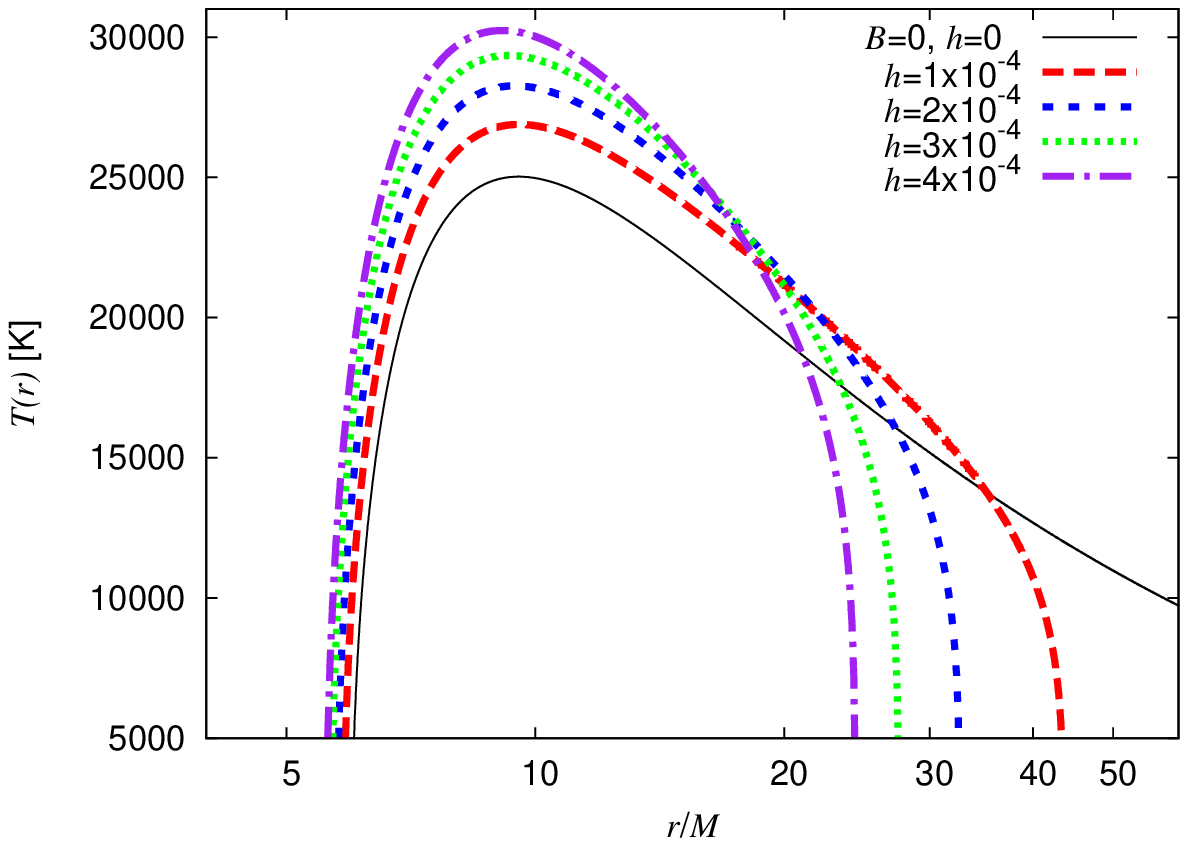} %
\includegraphics[width=8.7cm]{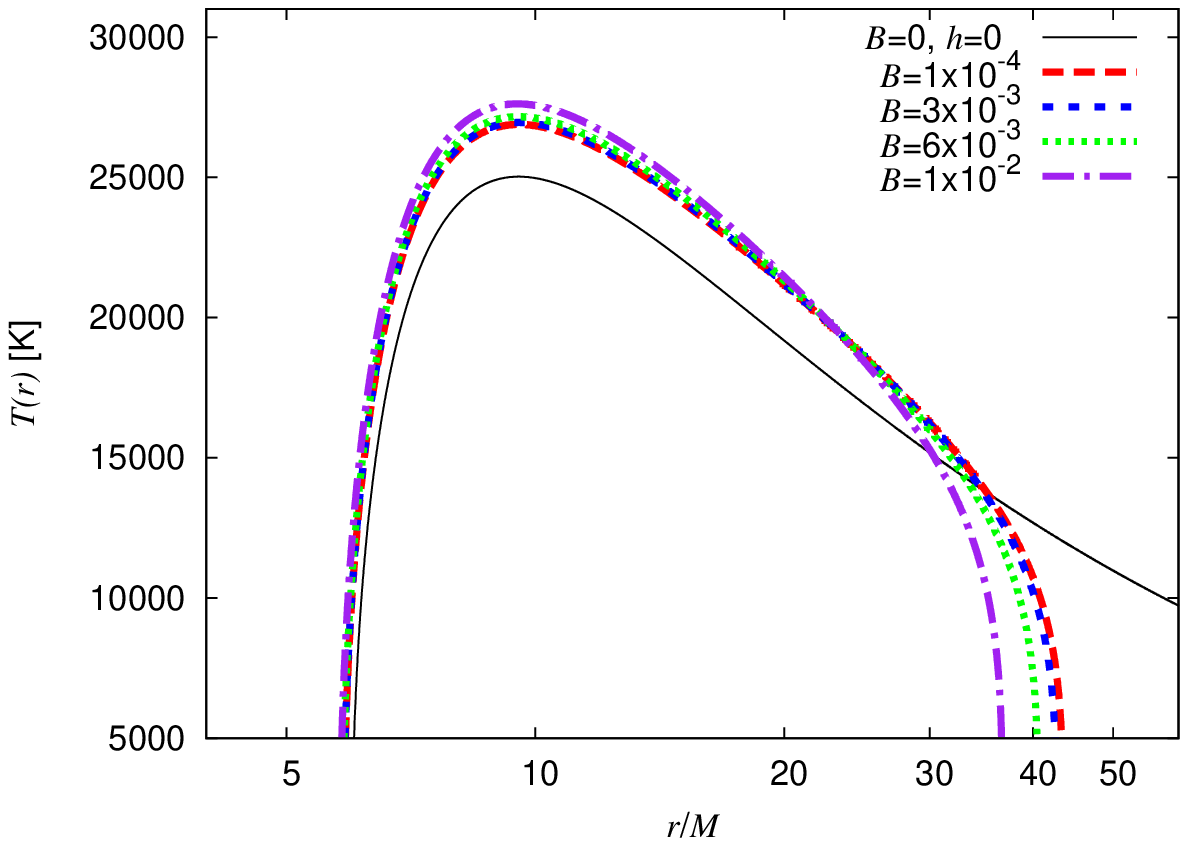}
\caption{The temperature profiles of an accretion disk around a perturbed
Schwarzschild black hole of total mass $M=2\times10^{6}M_{\bigodot}$. The
accretion rate is $2.5\times10^{-6}M_{\bigodot}$/yr. The solid line is the
temperature profile for a Schwarzschild black hole with the same total mass (%
$B=0$ and $K=0$). On the left plot $B$ is set to $10^{-4}M^{-1}$ and the
parameter $h$ is running, while on the right $h=10^{-4}M^{-2}$ is fixed and
different values of $B$ are taken.}
\label{T_Fig}
\end{figure}

Similar signatures can be recognized in the radial profiles of the disk
temperature, shown in Fig \ref{T_Fig} for the same parameter set of $B$ and
$K$ (or $h$).

\subsection{Disk spectrum}

\begin{figure}[tbp]
\includegraphics[width=8.7cm]{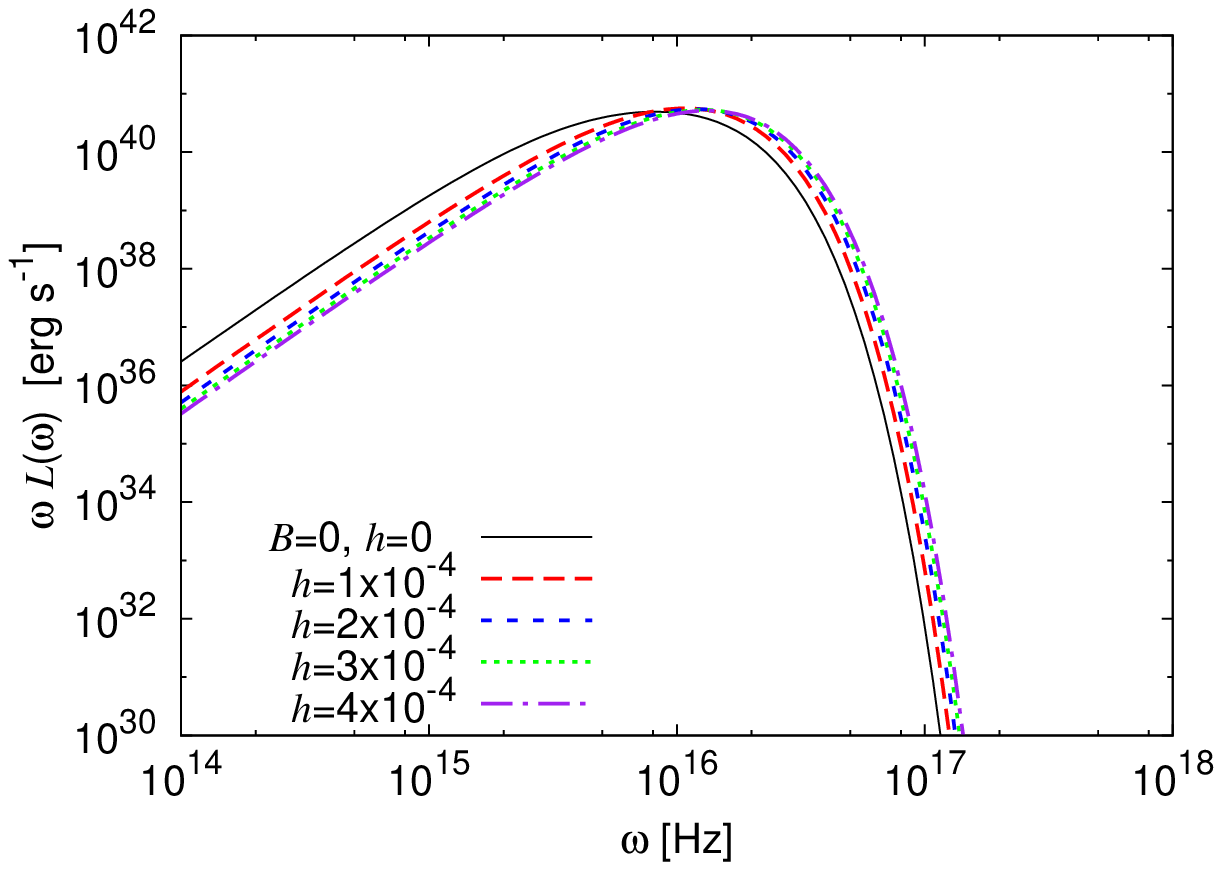} %
\includegraphics[width=8.7cm]{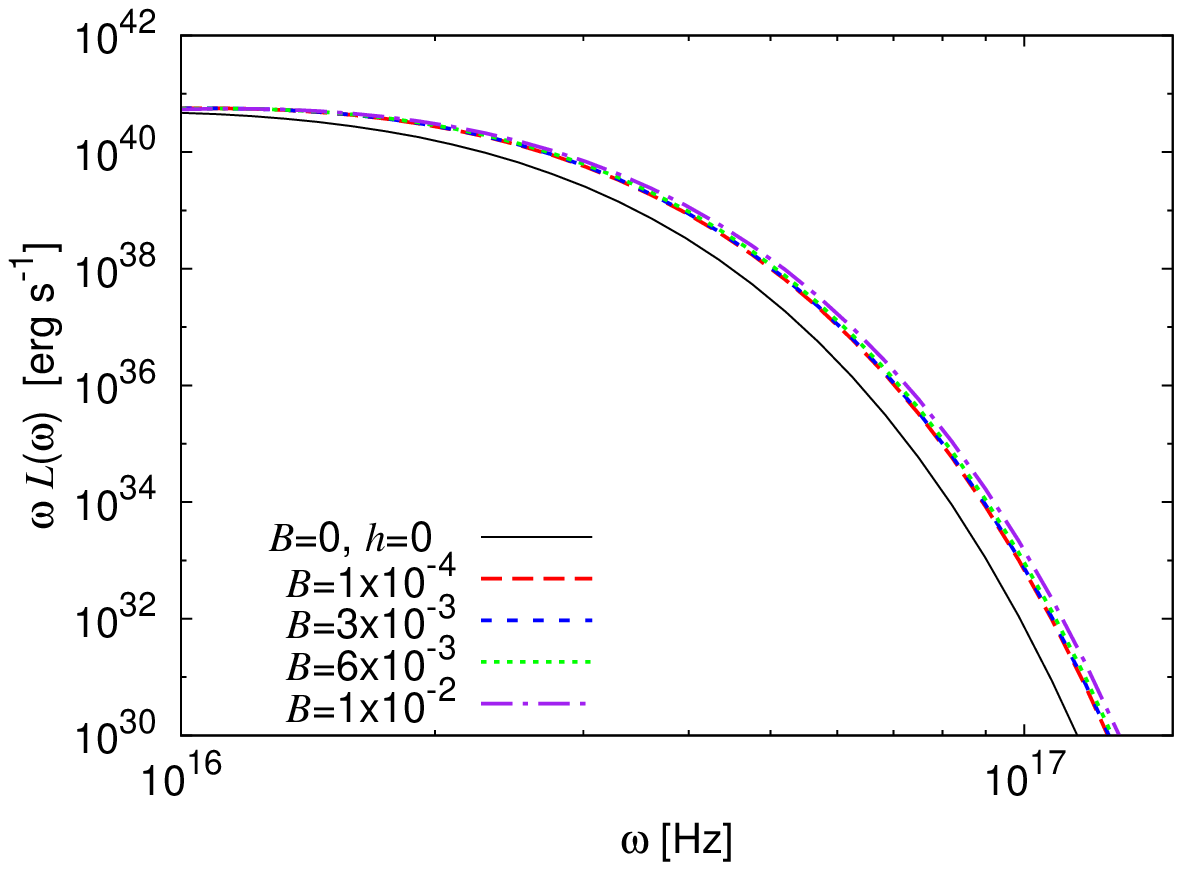}
\caption{The disk spectra for a perturbed Schwarzschild black hole of total
mass $M=2\times10^{6}M_{\bigodot}$. The accretion rate is $%
2.5\times10^{-6}M_{\bigodot}$/yr. The solid line is the disk spectrum for a
Schwarzschild black hole with the same total mass ($B=0$ and $K=0$). On the
left plot $B$ is set to $10^{-4}M^{-1}$ and the parameter $h$ is running,
while on the right $h=10^{-4}M^{-2}$ is fixed and different values of $B$
are taken.}
\label{L_Fig}
\end{figure}

The spectrum of the disk is derived from Eq.~(\ref{L}) and represented with
the same values of the perturbations as for the other plots. The
characteristic shape of the spectra on Fig \ref{L_Fig} shows a uniform
increase at low frequencies (on a logarithmic scale) followed by a sharp
decrease at high frequencies, ending in a cut-off at $\sim 10^{16}$ Hz.
Moreover, we note that in the presence of perturbations the spectrum is
blue-shifted in comparison with the Schwarzschild case. The shift of the
spectrum towards higher energies indicates that besides the accreted
mass-energy also some magnetic field energy is converted into radiation.

Finally, we give the conversion efficiency $\epsilon$ of the accreting mass
into radiation in the perturbed system for the different values of the
parameters $B$ and $K$ employed earlier in our analysis. In Table~\ref%
{eps_Tab} we give both the marginally stable orbit, at which the specific
energy is evaluated in the calculation of $\epsilon$ given in Eq. (\ref%
{epsilon}), and the efficiency for the indicated values of the parameters.
As the perturbation parameters increase, both $r_{ms}$ and the efficiency of
energy generation by accretion decrease.

\begin{table}[tbp]
\begin{centering}\begin{tabular}{llcccc}
$B[M^{-1}]$ & $h[M^{-2}]$ & $r_{ms}[M]$ & $\epsilon$ & &
\tabularnewline \hline 0 & 0 & 6.00 & 0.0572 & & \tabularnewline
\hline $10^{-4}$ & $10^{-4}$ & 5.86 & 0.0537 & & \tabularnewline &
$2\times10^{-4}$ & 5.75 & 0.0503 & & \tabularnewline &
$3\times10^{-4}$ & 5.67 & 0.0469 & & \tabularnewline &
$4\times10^{-4}$ & 5.59 & 0.0437 & & \tabularnewline \hline
$2\times10^{-3}$ & $10^{-4}$ & 5.85 & 0.0532 & & \tabularnewline
$8\times10^{-3}$ & & 5.83 & 0.0526 & & \tabularnewline $10^{-2}$ & &
5.81 & 0.0520 & & \tabularnewline & & & & & \tabularnewline
\end{tabular}\par\end{centering}
\caption{The marginally stable orbit and the conversion efficiency $\protect%
\epsilon$ of the magnetically perturbed Schwarzschild black hole for
different parameters $B$ and $K$ (or $h$).}
\label{eps_Tab}
\end{table}

\section{Concluding Remarks}

Astronomical observations of the accretion disks rotating around black holes
can provide both the spatial distribution (if the disk morphology is
resolved) and the spectral energy distribution of the thermal radiation
emitted by the disk. The radial flux profile and spectrum in the standard
thin disk model can in turn be calculated for various types of compact
central bodies with and without magnetosphere. Then a convenient way to
determine the mass and the spin of the central black hole is to fit the flux
profile and the spectrum derived from the simple disk model on the
observational data. For static black holes, the analysis of the deviations
of the disk radiation from the Schwarzschild case could indicate the
presence of a magnetic field.

In this paper we have discussed the mass accretion process in the region of
the Preston-Poisson space-time representing a Schwarzschild black hole
perturbed by a weak magnetic field (which is however asymptotically uniform)
and a distant tidal structure. For this we have (a) determined the region
where this interpretation holds; (b) corrected the dynamical equations of
test particles valid in the equatorial plane; and (c) applied the
hydrodynamic approximation for the orbiting plasma.

The study of the perturbations included in the accretion process showed that
(i) the thin disk model can be approximately applied until the radius where
the perturbed Schwarzschild black hole interpretation holds; (ii) the
accretion disk shrinks and the marginally stable orbit shifts towards the
black hole with the perturbation; (iii) the intensity of the radiation from
the accretion disk increases, while the radius where the radiation is
maximal remains unchanged; (iv) the spectrum is slightly blue-shifted; and
finally (v) the conversion efficiency of accreting mass into radiation is
decreased by both the magnetic and the tidal perturbations.

\begin{figure}[tbp]
\includegraphics[width=7cm]{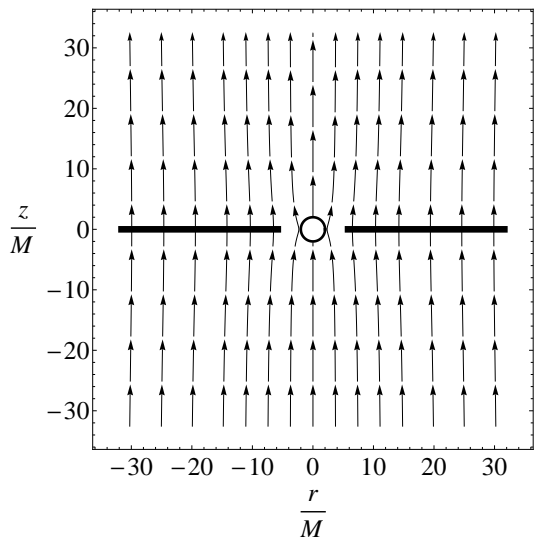}
\caption{The black hole horizon, the thin accretion disk (between $%
r_{ms}=5.8M$ and $r_{1}=31M$) and the magnetic field topology (for
$B=10^{-2}M^{-1}$). } \label{MagneticField2}
\end{figure}

We represent the system under discussion on Fig \ref{MagneticField2}.
Although the topology of magnetospheres around black holes is likely to be
more complicated than the simple model considered here, the blue shifted
disk spectrum indicating that some of the magnetic field energy also
contributes to the radiation may be a generic feature, signaling the
presence of a magnetic field. This conjecture is supported by the recent
finding that symbiotic systems of black holes in fast rotation, accretion
disk, jets and magnetic fields have a very similar magnetic field topology
to the one represented on Fig \ref{MagneticField2}, consisting of open
field lines only \cite{KGB}.

\section{Acknowledgements}

We thank Sergei Winitzki for interactions in the early stages of this
work. L\'{A}G is grateful to Tiberiu Harko for hospitality during his visit at the University of Hong Kong. L\'{A}G was partially supported by COST Action MP0905 "Black Holes in a Violent Universe". MV was supported by OTKA grant no. NI68228.

\appendix

\section{The curvature scalars on the horizon\label{curvature}}

In this Appendix we give the curvature scalars of the Preston-Poisson metric
(\ref{PPEdFink}). Throughout the computations (except otherwise stated) the
coordinates (\ref{t})-(\ref{rbar}) of Konoplya are used. The results are
valid up to $(B^{2},K)$ order. We found that:

(i) {The perturbations are such that the Ricci scalar vanishes to }$%
(B^{2},K) ${\ order}, $R=0$.

(ii) {With the use of light-cone gauge coordinates the Kretschmann scalar $%
\mathcal{K}=R_{abcd}R^{abcd}$ is
\begin{equation}
\mathcal{K}=\frac{48M}{r^{6}}\left[ M+Kr^{3}\left( 2-3\sin ^{2}\theta
\right) \right] ~,  \label{KPP}
\end{equation}%
which on the horizon (\ref{rH}) becomes
\begin{equation}
\mathcal{K}_{|r=r_{H}}=\frac{3}{4M^{4}}\left[ 1+16M^{2}K-4M^{2}(B^{2}+6K)%
\sin ^{2}\theta \right] ~.
\end{equation}%
Since the Ricci scalar vanishes the contraction of the Weyl tensor gives the
same expression, $C_{abcd}C^{abcd}=\mathcal{K}$. }

The Kretschmann scalar calculated in the Konoplya coordinates (\ref{t})-(\ref%
{rbar}) is
\begin{eqnarray}
\mathcal{K} &=&\frac{48M}{\bar{r}^{6}}\{M+2MB^{2}\bar{r}(2M-\bar{r})-2K\bar{r%
}(4M^{2}-2M\bar{r}-\bar{r}^{2})  \notag \\
&&+\bar{r}[3K(4M^{2}-2M\bar{r}-\bar{r}^{2})-2MB^{2}(3M-\bar{r})]\sin
^{2}\theta \}\ ,
\end{eqnarray}%
which agrees with (\ref{KPP}) after the coordinate transformation (\ref{rbar}%
).

(iii) {The contraction of the Weyl tensor with the Killing vectors $%
t^{a}=(1,0,0,0)$ and $\phi ^{a}=(0,0,0,1)$ is
\begin{eqnarray}
C_{abcd}t^{a}\phi ^{b}t^{c}\phi ^{d}=C_{0303}&=& \frac{(2M-\bar{r})\sin
^{2}\theta }{3\bar{r}^{2}}\{3M + [K(4M^2 - 4M\bar{r} + 3\bar{r}^2) - 2MB^2(M
- \bar{r})]\bar{r}  \notag \\
&& + 3M[2K(M^2 - M\bar{r} - \bar{r}^2) - B^2(M^2 - M\bar{r} + \bar{r}%
^2)]\sin^2\theta\}~.
\end{eqnarray}%
This quantity vanishes on the horizon. The contraction of the Killing
vectors with the Riemann tensor gives the same result since $%
R_{0303}=C_{0303}$. }

(iv) {The second order scalar invariants of the Riemann tensor are
\begin{eqnarray}
R_{\ abcd}^{\ast }R^{abcd} &=& {}^{\ast }R_{abcd}R^{abcd} = 0\ ,  \notag \\
{}^{\ast }R_{\ abcd}^{\ast }R^{abcd} &=& -\frac{16M}{\bar{r}^{10}\sin
^{2}\theta } \{ 3M + 2[2B^2M(7M - 4\bar{r}) - K(28M^2 - 16M\bar{r} - 3\bar{r}%
^2)]\bar{r}  \notag \\
&&-3[2B^2M(7M - 3\bar{r}) - K(28M^2 - 16M\bar{r} - 3\bar{r}^2)]\bar{r}%
\sin^2\theta \}\ ,
\end{eqnarray}%
where $R_{\ abcd}^{\ast }=e_{cd}^{\ \ pq}R_{abpq}/2$, ${}^{\ast
}R_{abcd}=e_{ab}^{\ \ pq}R_{pqcd}/2$ and ${}^{\ast }R_{\ abcd}^{\ast
}=e_{ab}^{\ \ pq}e_{cd}^{\ \ rs}R_{pqrs}/4$, and }$e_{abcd}$ is the
antisymmetric Levi-Civita symbol. Similar contractions with the Weyl tensor
give identical results.

On the horizon, the Euler scalar becomes
\begin{eqnarray}
{}^{\ast }R_{\ abcd}^{\ast }R^{abcd} = -\frac{1}{64M^8\sin ^{2}\theta } \{ 3
- 4M^2[2(B^2 - 8K) + 3(B^2 + 8K)\sin^2\theta] \}\ .
\end{eqnarray}

In conclusion, as the Kretschmann scalar $\mathcal{K}$ and the scalar $%
{}^{\ast }R_{\ abcd}^{\ast }R^{abcd}$ exhibit a $\theta $-dependence on the
horizon, we conclude that despite the spherical shape of the horizon in the
Konoplya coordinates, it has acquired a quadrupolar deformation due to the
perturbing magnetic and tidal effects.

\section{The energy-momentum tensor\label{Tab}}

Since the coordinate transformation from the Eddington-Finkelstein type
coordinates $(v,r,\theta ,\varphi )$ to Konoplya coordinates \cite{Konoplya}
does not affect the angular variables $\theta$ and $\varphi$ the form of the
vector potential (\ref{Aa}) remains unchanged in the new coordinate system.
Then the nonvanishing mixed components of the energy momentum tensor can be
written in Konoplya coordinates as
\begin{eqnarray}  \label{mixedK}
T^t_{\ t}&=&\frac{B^2}{8\pi\bar{r}}(2M\sin^2\theta-\bar{r})\ ,  \notag \\
T^{\bar{r}}_{\ \bar{r}}&=&-\frac{B^2}{8\pi\bar{r}}(2M-\bar{r}-2(M-\bar{r}%
)\cos^2\theta)\ ,  \notag \\
T^{\bar{r}}_{\ \theta}&=&-\frac{B^2}{4\pi}(2M-\bar{r})\cos\theta\sin\theta\ ,
\notag \\
T^{\theta}_{\ \bar{r}}&=&\frac{B^2}{4\pi\bar{r}} \cos\theta\sin\theta\ ,
\notag \\
T^\theta_{\ \theta}&=&\frac{B^2}{8\pi\bar{r}}(2M-\bar{r}-2(M-\bar{r}%
)\cos^2\theta)\ ,  \notag \\
T^\varphi_{\ \varphi}&=&-\frac{B^2}{8\pi\bar{r}}(2M\sin^2\theta-\bar{r})\ .
\end{eqnarray}
Transforming these tensor components from the Konoplya coordinates to the
coordinate system $x^{a^\prime}=(t,\bar{r},z=\bar{r}\cos\theta,\varphi)$
adapted to the equatorial plane, we obtain
\begin{eqnarray}
T^t_{\ t}=T^{\ t}_t&=&\frac{B^2}{8\pi\bar{r}^3}[2M(\bar{r}^2-z^2)-\bar{r}%
^3]\ ,  \notag \\
T^{\bar{r}}_{\ \bar{r}}=T^{\ \bar{r}}_{\bar{r}} &=&\frac{B^2}{8\pi\bar{r}^3}[%
\bar{r}^3-2M(\bar{r}^2+z^2)]\ ,  \notag \\
T^{\bar{r}}_{\ z}=T^{\ \bar{r}}_{z}&=&\frac{B^2}{4\pi\bar{r}^2}(2M-\bar{r}%
)z\ ,  \notag \\
T^{z}_{\ \bar{r}}=T^{\ z}_{\bar{r}}&=&-\frac{B^2}{2\pi\bar{r}^2}Mz\ ,  \notag
\\
T^z_{\ z}=T^{\ z}_z&=&\frac{B^2}{8\pi\bar{r}^3}[2M(\bar{r}^2+z^2)-\bar{r}%
^3]\ ,  \notag \\
T^\varphi_{\ \varphi}=T^{\ \varphi}_{\varphi}&=&\frac{B^2}{8\pi\bar{r}^3}%
[2M(z^2-\bar{r}^2)+\bar{r}^3]\ .
\end{eqnarray}
Since the components $T^r_t$, $T^z_t$ $T^r_{\varphi}$ and $T^z_{\varphi}$
vanish identically, they do not appear in the expressions $E^a=-T^a_{\ b}t^b$
and $J^a=T^a_{\ b}{\varphi}^b$ of the energy and angular momentum flux
4-vectors and, in turn, do not give any contributions to the integrated laws
of energy and angular momentum conservation.


\begin{thebibliography}{99}
\bibitem{ShSu73} N. I. Shakura and R. A. Sunyaev, Astron. Astrophys. \textbf{%
24}, 33 (1973).

\bibitem{NoTh73} I. D. Novikov and K. S. Thorne, in Black Holes, ed. C.
DeWitt and B. DeWitt, New York: Gordon and Breach (1973).

\bibitem{BZ77} R. D. Blandford and R. L. Znajek, {Month. Not. Roy. Astr. Soc.%
} \textbf{179}, 433 (1977).

\bibitem{Li02} L. X. Li, Astron. Astrophys. \textbf{392}, 469 (2002).

\bibitem{WXL02} D. X. Wang, K. Xiao, and W. H. Lei, {Month. Not. Roy. Astr.
Soc.} \textbf{335}, 655 (2002).

\bibitem{Ca86} M. Camenzind, {Astron. \& Astroph.} \textbf{156}, 137,
\textbf{1 62}, 32 (1986).

\bibitem{TNTT90} M. Takahashi, S. Nitta, Y. Tamematsu, and A. Tomimatsu, {%
Astrophys. J.} \textbf{363}, 206 (1990).

\bibitem{JC} A. Janiuk, B. Czerny, Month. Not. Roy. Astr. Soc. In press (2011),
E-print: arXiv:1102.3257.

\bibitem{KGB} Z. Kov{\'{a}}cs, L. \'{A}. Gergely, and P. L. Biermann, {%
Month. Not. Roy. Astr. Soc., in press (2011), }E-print: arXiv:1007.4279.

\bibitem{Uzdensky} D. A: Uzdensky, Astrophys. J. \textbf{603}, 652 (2004).

\bibitem{PP} B. Preston and E. Poisson, Phys. Rev. D \textbf{74}, 064010
(2006). 

\bibitem{KucharSchw} K. Kucha\v{r}, Phys. Rev. D \textbf{50}, 3961 (1994).

\bibitem{SchMelvin} F. J. Ernst, J. Math. Phys. (N.Y.) \textbf{17}, 54
(1976).

W. A. Hiscock, J. Math. Phys. (N.Y.) \textbf{22}, 1828 (1981).

F. J. Ernst and W. J. Wild, J. Math. Phys. (N.Y.) \textbf{17}, 182 (1976).

\bibitem{HH} S. W. Hawking and J. B. Hartle, Commun. Math. Phys. \textbf{27}%
, 283 (1972).

\bibitem{Konoplya} R. A. Konoplya, Phys. Rev. D \textbf{74}, 124015 (2006).

\bibitem{Wald} R. M. Wald, Phys. Rev. D \textbf{10}, 1680 (1974).

\bibitem{PaTh74} D. N. Page and K. S. Thorne, Astrophys. J. \textbf{191},
499 (1974).

\bibitem{Th74} K. S. Thorne, Astrophys. J. \textbf{191}, 507 (1974).
\end{thebibliography}
\end{document}